\documentclass[showpacs,amssymb,twocolumn,aps,nofootinbib]{revtex4}
\usepackage{amsmath}
\usepackage{amstext}
\usepackage{amsopn}
\usepackage{amsfonts}
\usepackage{amssymb}
\usepackage{bbm}
\usepackage{accents}
\usepackage{empheq}
\usepackage{graphicx}
\usepackage{epsf}
\usepackage{graphics}
\usepackage[latin1]{inputenc}
\def\sc{\scriptstyle}

\begin{document}

\title{Infrared divergences, mass shell singularities and gauge dependence of the dynamical fermion mass}

\author{Ashok K Das$^{a,b}$, J. Frenkel$^{c}$ and C. Schubert$^{d}$\footnote{$\ $ e-mail: das@pas.rochester.edu,  jfrenkel@fma.if.usp.br, schubert@ifm.umich.mx}}
\affiliation{$^a$ Department of Physics and Astronomy, University of Rochester, Rochester, NY 14627-0171, USA}
\affiliation{$^b$ Saha Institute of Nuclear Physics, 1/AF Bidhannagar, Calcutta 700064, INDIA}
\affiliation{$^{c}$ Instituto de Física, Universidade de São Paulo, 05508-090, São Paulo, SP, BRAZIL}
\affiliation{$^{d}$ Institute for Physics, Michoacan University, C. P. 58040, Morelia, Michoacan, MEXICO}

\begin{abstract}
We study the behavior of the dynamical fermion mass when infrared divergences and mass shell singularities are present in a gauge theory. In particular, in the massive Schwinger model in covariant gauges we find that the pole of the fermion propagator is divergent and gauge dependent at one loop, but the leading singularities cancel in the quenched rainbow approximation. On the other hand, in physical gauges, we find that the dynamical fermion mass is finite and gauge independent at least up to one loop.
\end{abstract}


\maketitle

\begin{figure}[ht!]
\begin{center}
\includegraphics[scale=.8]{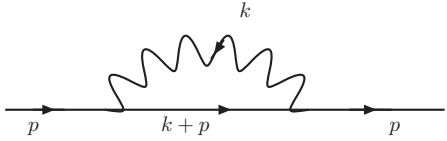}
\end{center}
\caption{Fermion self-energy in QED at one loop.}
\label{f1}
\end{figure}

The pole of the propagator for a particle defines the mass of the particle and, in a gauge theory, the pole is expected to be gauge independent \cite{tarrach, gray}. Intuitively it is clear that the physical mass of a particle should be gauge independent and quantitatively it can be argued as follows. Let us consider a fermion of tree level mass $m$ interacting with a photon field in a covariant gauge. Then the physical mass $m_{p}$ of the fermion is determined from the vanishing of the complete two point function $S^{-1} = p\!\!\!\slash - m - \Sigma (p,m) = 0$. Here $\Sigma (p,m)$ denotes the fermion self-energy to all orders. At one loop, for example, the fermion self-energy (see Fig. \ref{f1})
in $D$-dimensions  leads to the equation
\begin{equation}
\frac{\partial\Sigma^{(1)}}{\partial\xi} = e^{2} \left((p\!\!\!\slash - m) F_{1} + F_{2}(p\!\!\!\slash - m)\right),\label{1}
\end{equation}
where $\xi$ denotes the (covariant) gauge fixing parameter and
\begin{align}
F_{1} (p,m) & = \frac{1}{2}\int \frac{d^{\sc D}k}{(2\pi)^{\sc D}} \frac{1}{(k^{2})^{2}} \frac{1}{k\!\!\!\slash + p\!\!\!\slash - m}\, k\!\!\!\slash,\notag\\
F_{2} (p,m) & = \frac{1}{2}\int \frac{d^{\sc D}k}{(2\pi)^{\sc D}} \frac{1}{(k^{2})^{2}}\,  k\!\!\!\slash\, \frac{1}{k\!\!\!\slash + p\!\!\!\slash - m}.\label{2}
\end{align} 
Since to one loop order $m_{p}^{(1)} = m + O (e^{2})$ and the right hand side of \eqref{1} is already of order $e^{2}$, we can let $m\rightarrow m_{p}^{(1)}$ in the terms $(p\!\!\!\slash - m)$ on the right hand side which shows that the right hand side vanishes at the pole $p\!\!\!\slash = m_{p}^{(1)}$ if the functions $F_{1},F_{2}$ are well behaved. Furthermore, if we can Taylor expand the self-energy around the pole to write
\begin{equation}
\Sigma^{(1)} (p,m) = (m_{p}^{(1)} - m) + (p\!\!\!\slash - m_{p}^{(1)}) \left. (\Sigma^{(1)})' \right|_{p\!\!\!\slash =m_{p}^{(1)}} + \cdots,\label{3}
\end{equation}
only the first term survives at the pole. As a result, at the pole \eqref{1} leads to $\frac{\partial m_{p}^{(1)}}{\partial\xi} = 0$ showing that the pole of the propagator (mass of the particle) is gauge independent. Gauge independence of the pole at higher loops is best studied through an analysis of the Nielsen identity for the two point function which we will discuss later \cite{nielsen, brecken, das}.

It is clear from \eqref{2} that the functions $F_{1},F_{2}$ become infrared singular for space-time dimensions $D\leq 3$ so that the right hand side of \eqref{1} may not vanish in general. In addition, as we will see in lower dimensional models, mass shell singularities develop in the fermion self-energy so that a Taylor expansion (as in \eqref{3}) is not  possible and the gauge independence of the pole of the propagator can not be guaranteed. These are two possible difficulties that can invalidate the formal argument of gauge independence given above and when infrared divergences and/or mass singularities are present, the question of gauge independence of the pole  has to be examined anew. This is important since the effective action (and its imaginary part) for fermions interacting with a gauge field background depends on the fermion mass \cite{schwinger}  and being a physical quantity (the imaginary part is related to the vacuum decay rate), it cannot possibly be gauge dependent. 

Infrared singularities are well known to be a problem in lower dimensions \cite{swieca, wu}, but even in $4$-dimensional QCD, they can arise (in the fermion self-energy) at two loops and beyond because of gluon loops \cite{reinders}. Of course, intuitively it is clear that a strong infrared divergence signifies possible confinement of the fermion in the theory without any physical fermion asymptotic states. (As we know this is true even in Abelian theories in lower dimensions where the strong infrared behavior of the Coulomb force leads to confinement of fermions \cite{schwinger1, swieca, casher, jackiw, coleman}.) Nonetheless one can calculate the fermion mass for the perturbative asymptotic states as is done in the case of \lq\lq current" quark masses which are then used in various calculations and it is worth investigating whether such a calculation can lead to a gauge independent result. 

For these reasons we have chosen to study this question systematically. We find that in covariant gauges infrared divergences as well as mass shell singularities develop in the fermion self-energy at one loop when $D\leq 3$ invalidating the standard arguments for gauge independence of the pole of the propagator. However, we show that the leading divergences cancel when the fermion self-energy is summed to all orders in the quenched rainbow approximation \cite{maskawa, fukuda}. This signals the fact that the infrared divergences and mass shell singularities may not be present in a physical gauge such as the axial and Coulomb gauges. We carry out such a calculation for the $1+1$ dimensional massive QED and show explicitly that there are no infrared divergences in the self-energy up to two loops. We also find explicitly that the pole of the propagator is gauge independent at least up to one loop.

The one loop fermion self-energy in Fig. \ref{f1} can be calculated in a general covariant gauge in a straightforward manner and in dimensional regularization it has a closed form expression in terms of hypergeometric functions. However, to see the singularity structure, we note that near the mass shell $p^{2}\rightarrow m^{2}$ it behaves like
\begin{align}
&\Sigma^{(1)} (p) = - \frac{1}{2^{\sc D}\pi^{\frac{D}{2}}} \left(\frac{e}{m^{2-\frac{D}{2}}}\right)^{2}\Gamma(3- {\sc D})\notag\\
&\ \times\Big[- m({\sc D}-1+\xi) \Big(1-\frac{p^{2}}{m^{2}}\Big)^{{\sc D}-3} \Gamma (\frac{\sc D}{2}-1)\notag\\
&\ + 2\xi p\!\!\!\slash \Big(1-\frac{p^{2}}{m^{2}}\Big)^{{\sc D}-3} \Gamma(\frac{\sc D}{2})\notag\\
&\ + \left((-2+({\sc D}+1)\xi) p\!\!\!\slash - ({\sc D}-1+\xi)m\right)\Big(1-\frac{p^{2}}{m^{2}}\Big)^{{\sc D}-2} \Gamma(\frac{\sc D}{2})\notag\\
&\ - \left(\xi p\!\!\!\slash - ({\sc D}-1+\xi)m\right) \frac{\Gamma(2- \frac{D}{2})}{\Gamma(4-{\sc D})}\Big],\label{4}
\end{align}
This shows that for $D>3$, there are the usual ultraviolet divergences (coming from $\Gamma (3-D)$ and $\Gamma (2-\frac{D}{2})$), but there is no mass shell singularity so that the self-energy can have a proper Taylor expansion as in \eqref{3}. On the other hand, when $D\leq 3$ there are not only infrared divergences (the gamma function singularities in these cases can be identified with infrared divergences) but also mass shell singularities which show that a Taylor expansion of the self-energy is not possible. Explicitly, for example, we see from \eqref{4} that for $D = 2+ 2\epsilon$ the leading divergence of the self-energy has the form
\begin{equation}
\Sigma_{\sc L}^{(1)} (p) = \frac{e^{2}}{4\pi m} (1+\xi)\left(1- \frac{p^{2}}{m^{2}}\right)^{-1} \frac{1}{\epsilon}.\label{5}
\end{equation}
This is infrared divergent, has a mass shell singularity and is also explicitly gauge dependent. Although the leading divergence in \eqref{5} can be set to zero with the choice  $\xi=-1$, subleading divergences (which correspond to mass shell singularities)  persist. In fact, there is no value of $\xi$ which will make the self-energy completely well behaved. It is rather counterintuitive that the massive Schwinger model has an infrared divergence while the massless Schwinger model does not. This has its origin in the fact that the Dirac algebra in the self-energy in Fig. \ref{f1} leads to the numerator (say, in the Feynman gauge for illustration purposes) $ \gamma^{\mu} (p\!\!\!\slash + m)\gamma_{\mu} = (-2\epsilon p\!\!\!\slash + (2+2\epsilon) m)$. When $m\neq 0$, this numerator is finite in the limit $\epsilon\rightarrow 0$ and does not moderate the infrared divergence arising from the integral while when $m=0$, the numerator behaves as  $-2\epsilon p\!\!\!\slash$ and the $\epsilon$ in the numerator moderates the infrared behavior making it finite.

\begin{figure}[ht!]
\begin{center}
\includegraphics[scale=.8]{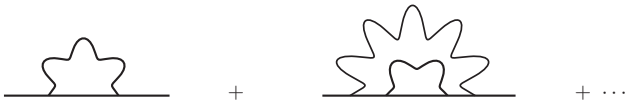}
\end{center}
\caption{Fermion self-energy diagrams in the quenched rainbow approximation.}
\label{f2}
\end{figure}

Since the one loop self-energy is divergent, it is worth asking if the higher loop corrections can cure the problem. A general analysis of this question is rather difficult. Instead we will show that when summed to all loops in the quenched rainbow approximation, the leading divergence in the fermion self-energy does indeed vanish. First let us recall that the overlapping divergence graphs, in general, have a softer infrared behavior \cite{taylor} so that in studying the leading divergence structure these can be neglected. Second, let us note here that when infrared divergences are present, normally including the polarization tensor in the photon propagator at higher loops improves the behavior. However, in the case of the massive Schwinger model, the one loop polarization tensor has the explicit form
\begin{equation}
\Pi^{\mu\nu} = \left(\eta^{\mu\nu} - \frac{p^{\mu}p^{\nu}}{p^{2}}\right)\Pi_{T} (p^{2}),\label{6}
\end{equation}
where
\begin{equation}
\Pi_{T} (p^{2}) = - \frac{e^{2}}{\pi}\left[1 - \int_{0}^{1} dx\,\frac{m^{2}}{m^{2} - x(1-x)p^{2}}\right].\label{7}
\end{equation}
It is clear from \eqref{7} that when the fermion mass vanishes $\Pi_{T} (p^{2}) = - \frac{e^{2}}{\pi}$ giving a mass to the photon in the (massless) Schwinger model. When $m\neq 0$, however, the function $\Pi_{T}(p^{2})$ is non-analytic. In fact, for $0\leq p^{2}\leq 4m^{2}$ we obtain from \eqref{7}
\begin{equation}
\Pi_{T} (p^{2}) = - \frac{e^{2}}{\pi}\left[1-\frac{\frac{4m^{2}}{p^{2}}}{\sqrt{\frac{4m^{2}}{p^{2}} -1}} \arctan \frac{1}{\sqrt{\frac{4m^{2}}{p^{2}} -1}}\right]\!\!,\label{8}
\end{equation}
while for $p^{2}\geq 4m^{2}$ we have from \eqref{7}
\begin{equation}
\Pi_{T} (p^{2}) = - \frac{e^{2}}{\pi}\left[1 - \frac{\frac{2m^{2}}{p^{2}}}{\sqrt{1 - \frac{4m^{2}}{p^{2}}}} \ln \frac{1-\sqrt{1 - \frac{4m^{2}}{p^{2}}}}{1+\sqrt{1 - \frac{4m^{2}}{p^{2}}}}\right]\!\!.\label{9}
\end{equation}
Equation \eqref{9} allows us to take the massless limit yielding the result for the Schwinger model. On the other hand, for a nonzero fermion mass, it is \eqref{8} which allows us to take the limit $p^{2}\rightarrow 0$ leading to
\begin{equation}
\Pi_{T} (p^{2})\xrightarrow{p^{2}\rightarrow 0} - \frac{e^{2}}{\pi} \left[1 - 1\right] = 0,\label{10}
\end{equation}
showing that a photon mass is not generated. (The absence of a \lq\lq Higgs" phenomena in the massive Schwinger model is already known from a different analysis \cite{jackiw, coleman}.) As a result, including vacuum polarization in the photon propagator in higher orders does not help and we restrict our analysis to the quenched rainbow approximation (see Fig. \ref{f2}).

The leading divergence terms in the self-energy diagrams in Fig. \ref{f2} for the massive Schwinger model can be calculated in a straightforward manner and the contribution at the $(n+1)$th loop turns out to be recursively related to the contribution at the $n$th loop in a simple manner
\begin{equation}
\Sigma_{L}^{(n+1)} (p) = \Sigma_{L}^{(n)} (p) \Sigma_{L}^{(1)} (p) \frac{1}{p\!\!\!\slash - m}\Big|_{p\!\!\!\slash\rightarrow m}.\label{11}
\end{equation}
This simple recursive behavior of the leading divergences in the quenched rainbow approximation allows us to sum the self-energy to all orders (in this approximation) and yields
\begin{align}
\Sigma_{L} & = \sum\limits_{n=0}^{\infty} \Sigma_{L}^{(n+1)} = \Sigma_{L}^{(1)} \sum\limits_{n=0}^{\infty} \left(\Sigma_{L}^{(1)} \frac{1}{p\!\!\!\slash - m}\right)^{n}\Big|_{p\!\!\!\slash\rightarrow m}\notag\\
& = \Sigma_{L}^{(1)} \frac{1}{p\!\!\!\slash - m - \Sigma_{L}^{(1)}}\, (p\!\!\!\slash - m)\Big|_{p\!\!\!\slash\rightarrow m} \rightarrow 0.\label{12}
\end{align}
This shows explicitly that the leading divergence vanishes in the self-energy when summed to all orders in the quenched rainbow approximation (for any value of $\xi$). This simple relation can also be seen in another way. We note that the leading term in the self-energy to all orders can be written as an integral equation (``gap" equation) of the form
\begin{align}
&\Sigma_{L}(p) = \Sigma_{L}^{(1)} (p) \notag\\
&+e^{2}\!\!\int \!\!\frac{d^{D}k}{(2\pi)^{D}} \gamma^{\mu}\frac{1}{p\!\!\!\slash+k\!\!\!\slash - m} \Sigma_{L}(p+k) \frac{1}{p\!\!\!\slash+k\!\!\!\slash -m}\gamma^{\nu} D_{\mu\nu}(k).\label{12a}
\end{align}
The leading infrared divergent contribution comes from the region $k\rightarrow 0$ and as $p\!\!\!\slash\rightarrow m$ this equation is solved in this regime by $\Sigma_{L} (p) = -(p\!\!\!\slash - m)\rightarrow 0$. (We have explicitly seen the same behavior in scalar QED in $1+1$ dimensions in covariant gauges as well.) 

Of course, this does not say anything about the subleading divergences which, in principle, are much harder to calculate since they involve the overlapping graphs. However, the vanishing of the leading divergence (when summed) suggests that the fermion self-energy may be better behaved when calculated in a physical gauge and this is what we discuss next. This can be understood as follows. In the covariant gauge, the fermion self-energy receives contribution from the unphysical as well as the physical degrees of the photon and a resummation may be necessary to see the cancellation between the contributions from these degrees of freedom. However, having seen the cancellation in the covariant gauge, it is clear that since a physical gauge has only physical degrees of freedom, there is no room for cancellation any more and the result should already be well behaved at one loop. In fact physical gauges, such as the axial gauge, are known to tame the infrared divergence much better and if there is no infrared divergence divergence at one loop (as we will see shortly), summing the rainbow graphs would no longer be a meaningful approximation.

Let us consider the massive Schwinger model in the homogeneous axial gauge $n\cdot A = 0$ where $n^{\mu}$ denotes a fixed direction in space-time. In this case, the photon propagator has the form
\begin{equation}
iD_{\mu\nu} (p) = - \frac{i}{p^{2}}\left(\eta_{\mu\nu} - \frac{n_{\mu}p_{\nu} + n_{\nu}p_{\mu}}{n\cdot p} + \frac{n^{2} p_{\mu}p_{\nu}}{(n\cdot p)^{2}}\right),\label{13}
\end{equation}
and satisfies
\begin{equation}
n^{\mu}D_{\mu\nu} (p) = 0 = D_{\mu\nu} n^{\nu}.\label{14}
\end{equation}
The poles of the propagator in \eqref{13} at $n\cdot p=0$ are conventionally handled through the PV (principal value)  prescription \cite{kummer, frenkel} which is known to soften the infrared behavior. It is also known that the PV prescription may be problematic \cite{leibbrandt} for time-like ($n^{2} > 0$) and light-like ($n^{2}=0$) axial gauges. For this reason we restrict ourselves to the class of space-like ($n^{2} < 0$) axial gauges. It is worth noting here that in $D=2$, the space-like axial gauges also include the Coulomb gauge. This can be seen in two different (but equivalent) ways. First, we note that if we choose the gauge $A_{1}=0$ or $n^{\mu} = (0,1)$, the only nontrivial component of the propagator in \eqref{13} is given by
\begin{equation}
iD_{00} (p) = \frac{i}{p_{1}^{2}},\label{15}
\end{equation}
which coincides with the propagator in the Coulomb gauge. Alternatively, we note that even after we impose the Coulomb gauge $\partial_{1}A_{1}=0$ which implies that $A_{1} = A_{1} (t)$, there is a residual gauge symmetry preserving this gauge choice, namely,
\begin{equation}
A_{1}\rightarrow A_{1}'  = A_{1} (t) + \partial_{1}\Lambda (t,x),\quad \partial_{1}A_{1}' = 0.\label{16}
\end{equation}
This determines $\Lambda (t,x) = x \alpha (t) + \beta (t)$ where $\alpha(t),\beta(t)$ are arbitrary functions of $t$. This leads to (see \eqref{16})
\begin{equation}
A_{1}'  = A_{1} (t) + \alpha (t),\label{17}
\end{equation}
and shows that if we choose $\alpha (t) = - A_{1}(t), A_{1}' = 0$ which corresponds to the axial gauge. Coulomb gauge is known to be notoriously singular in $D=2$. However, in considering this as belonging to the class of axial gauges in $D=2$, it inherits the PV prescription which gives it an improved behavior.

In $D=2$ the two vectors
\begin{equation}
n^{\mu}, n_{\sc T}^{\mu} = \epsilon^{\mu\nu} n_{\nu},\label{18}
\end{equation}
define a complete basis and, therefore, any vector (including the Dirac matrices $\gamma^{\mu}$) can be expanded in this basis. Using this the calculation of the fermion self-energy at one loop (see Fig. \ref{f1}) can be carried out in a manifestly covariant manner. The PV prescription indeed makes the integral infrared finite and we obtain
\begin{widetext}
\begin{equation}
\Sigma^{(1)} (p,m,n) = -\frac{e^{2}}{2\pi m} + \frac{e^{2}}{4\pi m (n\cdot p)} \left((p\!\!\!\slash - m) n\cdot \gamma + n\cdot\gamma (p\!\!\!\slash - m)\right)\left[1 + (\Omega^{2}-1) \left(1 - \frac{\Omega}{2}\,\ln \frac{1+\Omega}{1-\Omega}\right)\right],\label{19}
\end{equation}
\end{widetext}
where we have defined 
\begin{equation}
\Omega = \frac{n\cdot p}{\sqrt{(n\cdot p)^{2} - m^{2}n^{2}}}.\label{20}
\end{equation}
The seemingly divergent second term in \eqref{19}, in fact, vanishes as $n\cdot p\rightarrow 0$ so that the self-energy is completely well behaved. We also note that the complete gauge dependence ($n^{\mu}$ dependence) of the self-energy is contained in the second term which vanishes as $p\!\!\!\slash\rightarrow m$ so that the one loop correction to the pole of the fermion propagator is gauge independent. This calculation can also be extended to two loops. The axial gauge calculations are, in general, extremely difficult at two loops and beyond. We have not yet been able to evaluate the two loop self-energy explicitly. However, analyzing the integrand at this order (including overlapping graphs) we find that the integral appears to be infrared finite. But, without an explicit evaluation of the self-energy at two loops, it is not yet clear whether the pole of the propagator will be gauge independent at this order as well.

As we have mentioned earlier, the gauge dependence of the pole of the propagator is best analyzed using Nielsen identities which can be derived using the BRST symmetry \cite{brst} of the theory. Let us briefly sketch here how it is derived in a homogeneous axial gauge. The gauge fixed QED Lagrangian density (including free ghosts) in any dimension is given by 
\begin{align}
{\cal L} & = - \frac{1}{4} F_{\mu\nu} F^{\mu\nu} + \bar{\psi}\left(\gamma^{\mu} (i\partial_{\mu} - e A_{\mu}) - m\right)\psi\notag\\
&\qquad + F n\cdot A + \bar{c} n\cdot\partial c,\label{21}
\end{align}
where $F$ denotes an auxiliary field introduced for gauge fixing. We can introduce the Lagrangian density of the  standard sources needed to study BRST (Ward) identities. However, in order to study the gauge dependence of the fermion two point function we need to introduce an additional source $H^{\mu}$ (fermionic with ghost quantum number) for the composite field $(\bar{c}A_{\mu})$. Thus the complete Lagrangian density for the sources is given by
\begin{align}
{\cal L}_{\rm source} & = J^{\mu}A_{\mu} + JF + i (\bar{\psi}\chi - \bar{\chi}\psi) + i (\bar{c}\eta - \bar{\eta} c)\notag\\
& \quad + ie (\bar{M} c\psi - \bar{\psi} c M) + H^{\mu} \bar{c}A_{\mu},\label{22}
\end{align}
\begin{widetext}
\begin{figure}[ht!]
\begin{center}
\includegraphics[scale=.8]{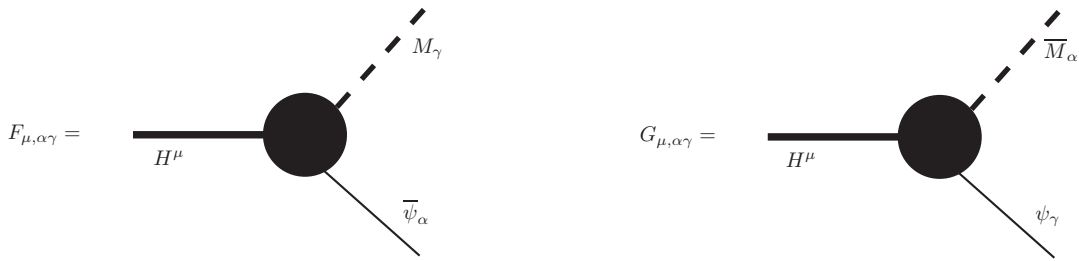}
\caption{The two three point vertices $F_{\mu}, G_{\mu}$.}
\label{f3}
\end{center}
\end{figure}
\end{widetext}
where $M, \bar{M}$ denote respectively the sources (bosonic spinors with anti-ghost/ghost quantum numbers) for the composite BRST variations of the fields $\bar{\psi}$ and $\psi$. With these it is straightforward to derive the identity describing the gauge dependence of the fermion two point function
\begin{equation}
\frac{\partial S^{-1}_{\alpha\beta} (p)}{\partial n^{\mu}} = F_{\mu, \alpha\gamma} (0, p, -p) S^{-1}_{\gamma\beta} (p) + S^{-1}_{\alpha\gamma} (p) G_{\mu, \gamma\beta} (0, -p, p),\label{23}
\end{equation}
where $\alpha,\beta,\gamma$ denote the Dirac indices and $F_{\mu}, G_{\mu}$ denote the three point vertices shown in Fig. \ref{f3}. Using the form of the photon propagator in \eqref{13}, it is straightforward to check that
\begin{equation}
\frac{\partial D_{\mu\nu} (p)}{\partial n^{\lambda}} = - \frac{p_{\mu}}{n\cdot p} D_{\lambda\nu} (p) - \frac{p_{\nu}}{n\cdot p} D_{\mu\lambda} (p).\label{24}
\end{equation}
Furthermore, using this relation (Eq. \eqref{24}) we can explicitly check graphically up to two loops that the Nielsen identity \eqref{23} holds which is consistent with our earlier observation that there is no infrared divergence at least up to two loops.

The Nielsen identity \eqref{23} can be studied order by order in a loop expansion. We note from \eqref{22} that there are no tree level three point vertices corresponding to $F_{\mu,\alpha\gamma}, G_{\mu,\alpha\gamma}$ (see Fig. \ref{f3}). They can arise only through radiative corrections and this is consistent with our expectation that gauge dependence in the fermion mass can possibly arise at the loop level. The vertices $F_{\mu,\alpha\gamma}, G_{\mu,\alpha\gamma}$ indeed arise beginning with one loop. Therefore, in the identity \eqref{23} at one loop, the fermion two point functions on the right hand side have to be the ones at the tree level. As a result, the analysis simplifies and gauge independence of the fermion mass (pole of the propagator) at one loop follows in a simple manner. This is, of course, completely in agreement with our explicit evaluation discussed in \eqref{19}. At higher loops, however, the analysis is more involved than in covariant gauges. This is because in any non covariant gauge involving a distinct tensor structure (such as $n^{\mu}$), the structure of $S^{-1} (p)$ becomes quite distinct from that of the tree level two point function \cite{konetschny}. For example, in two dimensions in the axial gauge we can expand the fermion two point function at higher loops in a complete basis of Dirac matrices $\Gamma^{A} = (\mathbbm{1}, \gamma^{\mu}, \gamma_{5}), \mu=0,1$ as
\begin{equation}
S^{-1} (p,m,n) = a_{\sc A} \Gamma^{A} = a_{\sc I} \mathbbm{1} + a_{\mu} \gamma^{\mu} + a_{\sc P} \gamma_{5},\label{25}
\end{equation}
where the coefficients $a_{\sc A}$ are, in general Lorentz covariant functions of $(m, p^{\mu}, n^{\mu})$. (This has to be contrasted with the covariant gauges where the coefficients have to be functions of $(m, p^{\mu})$ which makes the two point function to be of the same form as in the tree level theory.) The dependence on a new structure makes it much harder to analyze the question of gauge independence of the pole. As a result, even though we know the Nielsen identity to hold at two loop level, we have not yet been able to conclude whether the pole of the propagator is gauge independent at that order. This is consistent with our earlier observation that even though there is no infrared divergence at two loops, we are unable to say if the pole is gauge independent without explicitly evaluating the self-energy at two loops. A general study of the analysis of gauge independence directly from the Nielsen identity is presently underway.

To summarize, we have shown explicitly that the presence of infrared divergences as well as mass shell singularities can invalidate the conventional arguments for the gauge independence of the pole of the fermion propagator. We have shown this explicitly in the case of the one loop fermion self-energy in the (two dimensional) massive Schwinger model in covariant gauges. We have also shown that the leading divergence adds up to zero when all the diagrams are summed in the quenched rainbow approximation. This suggests that the self-energy may be better behaved in a physical gauge. We have shown that this is indeed the case in the (space-like) axial gauges where the PV (principal value) prescription softens the infrared divergence and leads to a gauge independent pole of the propagator at one loop. We also find that the integrand (in the self-energy) leads to an infrared finite integral at two loops although we have not yet been able to explicitly evaluate the integral. Therefore, we are not yet able to say if the pole of the fermion propagator is gauge independent even at two loops. The same conclusions follow from an analysis of the Nielsen identity up to this order and this question is being further studied. If the axial gauge does allow a gauge independent definition of the dynamical fermion  mass in the massive Schwinger model, it would indeed be worth trying such an analysis in the calculation of the quark mass in QCD. Details of our calculations will be published elsewhere.

\bigskip

\noindent{\bf Acknowledgments}
\medskip

This work was supported in part by CNPq and FAPESP (Brazil).

\end{document}